\documentclass[12pt,english]{paper}
\usepackage[T1]{fontenc}
\usepackage[latin1]{inputenc}
\usepackage{graphicx}

\makeatletter

\makeatletter

\usepackage{lineno}

\makeatletter

\usepackage{geometry}

\makeatletter

\usepackage{epsfig}

\makeatother

\makeatother

\makeatother

\usepackage{babel}
\makeatother
\begin{document}

\title{Applying Bayesian Neural Network to Determine Neutrino Incoming Direction
in Reactor Neutrino Experiments and Supernova Explosion Location by
Scintillator Detectors}

\author{ Weiwei Xu$^a$, Ye Xu$^a$%
\thanks{Corresponding author, e-mail address: xuye76@nankai.edu.cn%
}, Yixiong Meng$^a$, Bin Wu$^a$ }

\maketitle
\begin{flushleft}
$^a$Department of Physics, Nankai University, Tianjin 300071, The
People's Republic of China
\par\end{flushleft}

\begin{abstract}
In the paper, it is discussed by using Monte-Carlo simulation that
the Bayesian Neural Network(BNN) is applied to determine neutrino
incoming direction in reactor neutrino experiments and supernova
explosion location by scintillator detectors. As a result, compared
to the method in Ref.\cite{key-1}, the uncertainty on the
measurement of the neutrino direction using BNN is significantly
improved. The uncertainty on the measurement of the reactor neutrino
direction is about 1.0$^\circ$ at the 68.3\% C.L., and the one in
the case of supernova neutrino is about 0.6$^\circ$ at the 68.3\%
C.L.. Compared to the method in Ref.\cite{key-1}, the uncertainty
attainable by using BNN reduces by a factor of about 20. And
compared to the Super-Kamiokande experiment(SK), it reduces by a
factor of about 8.
\end{abstract}
\begin{flushleft}
\par\end{flushleft}

\section{Introduction}
The location of a $\nu$ source is very important to study galactic
supernova explosion. The determination of neutrino incoming
direction can be used to locate a supernova, especially, if the
supernova is not optically visible. The method based on the inverse
$\beta$ decay, $\bar{\nu_{e}}+p\rightarrow e^{+}+n$, has been
discussed in the Ref.\cite{key-1}. The method can be applied to
determine a reactor neutrino direction and a supernova neutrino
direction. But the uncertainty of location of the $\nu$ source
attainable by using the method is not small enough and almost 2
times as large as that in the Super-Kamiokande experiment(SK). So we
try to apply the Bayesian neural network(BNN)\cite{key-2} to locate
$\nu$ sources in order to decrease the uncertainty on the
measurement of the neutrino incoming direction.
\par

BNN is an algorithm of the neural networks trained by Bayesian
statistics. It is not only a non-linear function as neural networks,
but also controls model complexity. So its flexibility makes it
possible to discover more general relationships in data than the
traditional statistical methods and its preferring simple models
make it possible to solve the over-fitting problem better than the
general neural networks\cite{key-3}. BNN has been used to particle
identification and event reconstruction in the experiments of the
high energy physics, such as Ref.\cite{key-4,key-5,key-6,key-7}.
\par
In this paper, it is discussed by using Monte-Carlo simulation that
the method of BNN is applied to determine neutrino incoming
direction in reactor neutrino experiments and supernova explosion
location by scintillator detectors.

\section{Regression with BNN\cite{key-2,key-6}}

The idea of BNN is to regard the process of training a neural
network as a Bayesian inference. Bayes' theorem is used to assign a
posterior density to each point, $\bar{\theta}$, in the parameter
space of the neural networks. Each point $\bar{\theta}$ denotes a
neural network. In the method of BNN, one performs a weighted
average over all points in the parameter space of the neural
network, that is, all neural networks. The methods make use of
training data \{($x_{1}$,$t_{1}$),
($x_{2}$,$t_{2}$),...,($x_{n}$,$t_{n}$)\}, where $t_{i}$ is the
known target value associated with data $x_{i}$, which has $P$
components if there are $P$ input values in the regression. That is
the set of data $x=$($x_{1}$,$x_{2}$,...,$x_{n}$) which corresponds
to the set of target $t=$($t_{1}$,$t_{2}$,...,$t_{n}$). The
posterior density assigned to the point $\bar{\theta}$, that is, to
a neural network, is given by Bayes' theorem

\begin{center}
\begin{equation}
p\left(\bar{\theta}\mid
x,t\right)=\frac{\mathit{p\left(x,t\mid\bar{\theta}\right)p\left(\bar{\theta}\right)}}{p\left(x,t\right)}=\frac{p\left(t\mid
x,\bar{\theta}\right)p\left(x\mid\bar{\theta}\right)p\left(\bar{\theta}\right)}{p\left(t\mid
x\right)p\left(x\right)}=\frac{\mathit{p\left(t\mid
x,\bar{\theta}\right)p\left(\bar{\theta}\right)}}{p\left(t\mid
x\right)}\end{equation}

\par\end{center}

\begin{flushleft}
where data $x$ do not depend on $\bar{\theta}$, so
$p\left(x\mid\theta\right)=p\left(x\right)$. We need the likelihood
$p\left(t\mid x,\bar{\theta}\right)$ and the prior density
$p\left(\bar{\theta}\right)$, in order to assign the posterior
density $p\left(\bar{\theta}\mid x,t\right)$to a neural network
defined by the point $\bar{\theta}$. $p\left(t\mid x\right)$ is
called evidence and plays the role of a normalizing constant, so we
ignore the evidence. That is,
\par\end{flushleft}

\begin{center}
\begin{equation}
Posterior\propto Likelihood\times Prior\end{equation}

\par\end{center}

\begin{flushleft}
We consider a class of neural networks defined by the function
\par\end{flushleft}

\begin{center}
\begin{equation}
y\left(x,\bar{\theta}\right)=b+{\textstyle {\displaystyle
\sum_{j=1}^{H}v_{j}sin\left(a_{j}+\sum_{i=1}^{P}u_{ij}x_{i}\right)}}\end{equation}
\par\end{center}

\begin{flushleft}
The neural networks have $P$ inputs, a single hidden layer of $H$
hidden nodes and one output. In the particular BNN described here,
each neural network has the same structure. The parameter $u_{ij}$
and $v_{j}$ are called the weights and $a_{j}$ and $b$ are called
the biases. Both sets of parameters are generally referred to
collectively as the weights of the BNN, $\bar{\theta}$.
$y\left(x,\bar{\theta}\right)$ is the predicted target value. We
assume that the noise on target values can be modeled by the
Gaussian distribution. So the likelihood of $n$ training events is
\par\end{flushleft}

\begin{center}
\begin{equation}
p\left(t\mid
x,\bar{\theta}\right)=\prod_{i=1}^{n}exp[-((t_{i}-y\left(x_{i},\bar{\theta}\right))^{2}/2\sigma^{2}]=exp[-\sum_{i=1}^{n}(t_{i}-y\left(x_{i},\bar{\theta}\right)/2\sigma^{2})]\end{equation}

\par\end{center}

\begin{flushleft}
where $t_{i}$ is the target value, and $\sigma$ is the standard
deviation of the noise. It has been assumed that the events are
independent with each other. Then, the likelihood of the predicted
target value is computed by Eq. (4).
\par\end{flushleft}

We get the likelihood, meanwhile we need the prior to compute the
posterior density. But the choice of prior is not obvious. However,
experience suggests a reasonable class is the priors of Gaussian
class centered at zero, which prefers smaller rather than larger
weights, because smaller weights yield smoother fits to data . In
the paper, a Gaussian prior is specified for each weight using the
BNN package of Radford Neal%
\footnote{R. M. Neal, \emph{Software for Flexible Bayesian Modeling
and Markov
Chain Sampling}, http://www.cs.utoronto.ca/\textasciitilde{}radford/fbm.software.html%
}. However, the variance for weights belonging to a given
group(either input-to-hidden weights($u_{ij}$), hidden
-biases($a_{j}$), hidden-to-output weights($v_{j}$) or
output-biases($b$)) is chosen to be the same: $\sigma_{u}^{2}$,
$\sigma_{a}^{2}$, $\sigma_{v}^{2}$, $\sigma_{b}^{2}$, respectively.
However, since we don't know, a priori, what these variances should
be, their values are allowed to vary over a large range, while
favoring small variances. This is done by assigning each variance a
gamma prior
\par

\begin{center}
\begin{equation}
p\left(z\right)=\left(\frac{\alpha}{\mu}\right)^{\alpha}\frac{z^{\alpha-1}e^{-z\frac{\alpha}{\mu}}}{\Gamma\left(\alpha\right)}\end{equation}

\par\end{center}

\begin{flushleft}
where $z=\sigma^{-2}$, and with the mean $\mu$ and shape parameter
$\alpha$ set to some fixed plausible values. The gamma prior is
referred to as a hyperprior and the parameter of the hyperprior is
called a hyperparameter.
\end{flushleft}
\par
Then, the posterior density, $p\left(\bar{\theta}\mid x,t\right)$,
is gotten according to Eqs. (2),(4) and the prior of Gaussian
distribution. Given an event with data $x'$, an estimate of the
target value is given by the weighted average
\par

\begin{center}
\begin{equation}
\bar{y}\left(x'|x,t\right)=\int
y\left(x',\bar{\theta}\right)p\left(\bar{\theta}\mid
x,t\right)d\bar{\theta}\end{equation}

\par\end{center}

\begin{flushleft}
Currently, the only way to perform the high dimensional integral in
Eq. (6) is to sample the density $p\left(\bar{\theta}\mid
x,t\right)$ with the Markov Chain Monte Carlo (MCMC)
method\cite{key-2,key-8,key-9,key-10}. In the MCMC method, one steps
through the $\bar{\theta}$ parameter space in such a way that points
are visited with a probability proportional to the posterior
density, $p\left(\bar{\theta}\mid x,t\right)$. Points where
$p\left(\bar{\theta}\mid x,t\right)$ is large will be visited more
often than points where $p\left(\bar{\theta}\mid x,t\right)$ is
small.
\par\end{flushleft}

\begin{flushleft}
Eq. (6) approximates the integral using the average
\par\end{flushleft}

\begin{center}
\begin{equation}
\bar{y}\left(x'\mid
x,t\right)\approx\frac{1}{L}\sum_{i=1}^{L}y\left(x',\bar{\theta_{i}}\right)\end{equation}

\par\end{center}

\begin{flushleft}
where $L$ is the number of points $\bar{\theta}$ sampled from
$p\left(\bar{\theta}\mid x,t\right)$. Each point $\bar{\theta}$
corresponds to a different neural network with the same structure.
So the average is an average over neural networks, and is closer to
the real value of $\bar{y}\left(x'\mid x,t\right)$, when $L$ is
sufficiently large.
\par\end{flushleft}

\section{Toy Detector and Simulation\cite{key-5}}

In the paper, a toy detector is designed to simulate the central
detector in the reactor neutrino experiment, such as Daya Bay
experiment\cite{key-11} and Double CHOOZ experiment\cite{key-12},
with CERN GEANT4 package\cite{key-13}. The toy detector consists of
three regions, and they are the Gd-doped liquid scintillator(Gd-LS
from now on), the normal liquid scintillator(LS from now on) and the
oil buffer, respectively. The toy detector of cylindrical shape like
the detector modules of Daya Bay experiment and Double CHOOZ
experiment is designed in the paper. The diameter of the Gd-LS
region is 2.4 meter, and its height is 2.6 meter. The thickness of
the LS region is 0.35 meter, and the thickness of the oil part is
0.40 meter. In the paper, the Gd-LS and LS are the same as the
scintillator adopted by the proposal of the CHOOZ
experiment\cite{key-14}. The 8-inch photomultiplier tubes (PMT from
now on) are mounted on the inside the oil region of the detector. A
total of 366 PMTs are arranged in 8 rings of 30 PMTs on the lateral
surface of the oil region, and in 5 rings of 24, 18, 12, 6, 3 PMTs
on the top and bottom caps.

\par
The response of the neutrino and background events deposited in the
toy detector is simulated with GEANT4. Although the physical
properties of the scintillator and the oil (their optical
attenuation length, refractive index and so on) are wave-length
dependent, only averages\cite{key-14} (such as the optical
attenuation length of Gd-LS with a uniform value is 8 meter and the
one of LS is 20 meter) are used in the detector simulation. The
program couldn't simulate the real detector response, but this won't
affect the result of the comparison between the BNN and the method
in the Ref.\cite{key-1}.

\section{Event Reconstruction\cite{key-5}}

The task of the event reconstruction in the reactor neutrino
experiments is to reconstruct the energy and the vertex of a signal.
The maximum likelihood method (MLD) is a standard algorithm of the
event reconstruction in the reactor neutrino experiments. The
likelihood is defined as the joint Poisson probability of observing
a measured distribution of photoelectrons over the all PMTs for
given ($E,\overrightarrow{x}$) coordinates in the detector. The
Ref.\cite{key-15} for the work of the CHOOZ experiment shows the
method of the reconstruction in detail.

In the paper, the event reconstruction with the MLD are performed in
the similar way with the CHOOZ experiment\cite{key-15}, but the
detector is different from the detector of the CHOOZ experiment, so
compared to Ref.\cite{key-15}, there are some different points in
the paper:

(1) The detector in the paper consists of three regions, so the path
length from a signal vertex to the PMTs consist of three parts, and
they are the path length in Gd-LS region, the one in LS region, and
the one in oil region, respectively.

(2) Considered that not all PMTs in the detector can receive
photoelectrons when a electron is deposited in the detector, the
$\chi^{2}$ equation is modified in the paper and different from the
one in the CHOOZ experiment, that is,
$\chi^{2}=\sum_{N_{j}=0}\bar{N_{j}}+\sum_{N_{j}\neq0}(\bar{N}_{j}-N_{j}+N_{j}log(\frac{N_{j}}{\bar{N_{j}}}))$,
where $N_{j}$ is the number of photoelectrons received by the j-th
PMT and $\bar{N_{j}}$ is the expected one for the j-th
PMT\cite{key-15}.

(3) $c_{E}\times N_{total}$ and the coordinates of the charge center
of gravity for the all visible photoelectrons from a signal are regarded
as the starting values for the fit parameters($E,\overrightarrow{x}$),
where $N_{total}$ is the total numbers of the visible photoelectrons
from a signal and $c_{E}$ is the proportionality constant of the
energy $E$, that is, $E=c_{E}\times N_{total}$. $c_{E}$ is obtained
through fitting $N_{total}$'s of the 1 MeV electron events, and is
$\frac{1}{235/MeV}$ in the paper.

\section{Monte-Carlo Sample}
\subsection{Monte-Carlo Sample for Reactor Neutrinos}
According to the anti-neutrino interaction in the detector of the
reactor neutrino experiments\cite{key-16}, the neutrino events from
the random direction and the particular direction,
(0.433,0.75,-0.5), are generated uniformly throughout GD-LS region
of the toy detector. Fig. 1 shows the four important physics
quantities of the Monte-Carlo reactor neutrino events and they are
$E_{e^+}, E_n,
$$\Delta$$t_{e^+n}$$, d_{e^+n}$, respectively. The selections of
the neutrino events are as follows:

\par
(1) Positron energy: 1.3 MeV < $E_{e^{+}}$ < 8 MeV;

(2) Neutron energy: 6 MeV < $E_{n}$ < 10 MeV;

(3) Neutron delay: 2 $\mu$s < $\Delta$$t_{e^{+}n}$ < 100 $\mu$s;

(4) Relative positron-neutron distance: $d_{e^{+}n}$ < 100 cm.

\par
10000 events from the random directions and 5000 events from
(0.433,0.75,-0.5) are selected according to the above criteria,
respectively. The events from the random direction are regarded as
the training sample of BNN, and the events from (0.433,0.75,-0.5)
are regarded as the test sample of BNN.

\subsection{Monte-Carlo sample for Supernova Neutrinos}
The neutrino events for the random direction and the particular
direction, (0.354,0.612,-0.707), are generated uniformly throughout
GD-LS region of a liquid scintillator detector with the same
geometry and the same target as the toy detector in the sec. 3,
according to the following supernova $\bar{\nu_e}$ energy
distribution\cite{key-1,key-17}:

\begin{equation}
\frac{dN}{dE}=C\frac{E^2}{1+e^{E/T}}
\end{equation}
\begin{flushleft}
with $T=3.3 MeV$ and the supernova is considered to be at $10 Kpc$.
The number of the fixed direction neutrino events, for a supernova
at $10 Kpc$, could be detected in a liquid scintillator experiment
with mass equal to that of SK\cite{key-1}. The events from the
random direction are regarded as the training sample of BNN, and the
events from (0.354,0.612,-0.707) are regarded as the test sample of
BNN. Fig. 2 shows the four important physics quantities of the
Monte-Carlo supernova neutrino events and they are $E_{e^+}, E_n,
$$\Delta$$t_{e^+n}$$, d_{e^+n}$, respectively.
\end{flushleft}

\section{Location of the neutrino source using the method in the Ref.\cite{key-1}}
The inverse-$\beta$ decay can be used to locate the neutrino source
in scintillator detector experiments. The method is based on the
neutron boost in the forward direction. And neutron retains a memory
of the neutrino source direction. The unit vector $\hat{X}_{e^+n}$,
having its origin at the positron reconstructed position and
pointing to the captured neutron position, is defined for each
neutrino event. The distribution of the projection of this vector
along the known neutrino direction is forward peaked , but its
r.m.s. value is not far from that of a flat
distribution($\sigma_{flat}=1/\sqrt{3}$). $\vec{p}$ is defined as
the average of vectors $\hat{X}_{e^+n}$, that is
\begin{center}
\begin{equation}
\vec{p}=\frac{1}{N}\sum\hat{X}_{e^+n}
\end{equation}
\par\end{center}
The measured neutrino direction is the direction of $\vec{p}$.

\par
The neutrino direction lies along the z axis is assumed to evaluate
the uncertainty in the direction of $\vec{p}$. From the central
limit theorem $\vec{p}$ follows that the distribution of the three
components is Gaussian with $\sigma=1/\sqrt{3N}$ centered at
(0,0,$|\vec{p}|$). Therefore, the uncertainty on the measurement of
the neutrino direction can be given as the cone around $\vec{p}$
which contains 68.3\% of the integral of this distribution.

\section{Location of the neutrino source using BNN}
In the paper, the x,y,z components of the neutrino incoming
direction are predicted by the three BNNs, respectively. The BNNs
have the input layer of 6 inputs, the single hidden layer of 15
nodes and the output layer of a output. Here we will explain the
case of predicting the x component of the neutrino incoming
direction in detail:
\par
 (1) The data format for the training sample is
{$d_i,f_i,E_{e^+},E_n,$$\Delta$$t_{e^+n}$$,d_{e^+n}, t_i$} (i=x),
where $d_i$ is the difference of $v_i$ and $n_i$ (i=x). $v_i$(i=x)
is the x components of the $\hat{X}_{e^+n}$ in the section 6.
$n_i$(i=x) is the x component of the
 known neutrino incoming direction ($\vec{n}$).
$f_i$(i=x) is the x component of the reconstructed positron
position. $d_i,f_i,E_{e^+},E_n,$$\Delta$$t_{e^+n}$$,d_{e^+n}$ are
used as inputs to a BNN, and $t_i$ is the known target. The target
can be obtained by Eq. 10. That is
 \begin{equation}
t_i=\frac{1}{1+exp(0.5v_i/n_i)}(i=x).
\end{equation}
 where
\par
 (2) The inputs of the test sample are similar with that of the train sample,
 but the $d_i$(i=x) is different from that of the training sample.
 The $\vec{p}$ obtained by the method in the
section 6 is substituted for the known neutrino incoming direction
in the process of computing $d_i$(i=x). The $tp_i$(i=x) is the
output of the BNN, that is, it is the predicted value using the BNN.
We make use of the $tp_i$ value to compute the x component of
neutrino incoming direction via the following equation(In fact, Eq.
11 is the inverse-function of Eq. 10.):
\begin{equation}
m_i=\frac{0.5v_i}{ln(1/tp_i-1)}(i=x),
\end{equation}
where $v_i$(i=x) is the x
 component of the $\hat{X}_{e^+n}$. $m_i$(i=x) is just the x component of the direction vector
($\vec{m}$) predicted by the BNN.
\par
A Markov chain of neural networks is generated using the BNN package
of Radford Neal, with the training sample, in the process of
predicting the x component of neutrino incoming direction by using
the BNN. One thousand iterations, of twenty MCMC steps each, are
used in the paper. The neural network parameters are stored after
each iteration, since the correlation between adjacent steps is very
high. That is, the points in neural network parameter space are
saved to lessen the correlation after twenty steps. It is also
necessary to discard the initial part of the Markov chain because
the correlation between the initial point of the chain and the
points of the part is very high. The initial three hundred
iterations are discarded in the paper.
\par
Certainly, the y,z components of the $\vec{m}$ are obtained in the
same method, if only i=y,z, respectively. Here $\vec{L}$ is defined
as the unit vector of the $\vec{m}$ predicted by the BNNs for each
event in the test sample. We can also define the direction $\vec{q}$
as the average of the unit direction vectors predicted by the BNNs
in the same way as the section 6. That is
\begin{equation}
\vec{q}=\frac{1}{N}\sum\vec{L}.
\end{equation}
The $\vec{q}$ is just the neutrino incoming direction predicted by
the BNNs. The uncertainty in this value is evaluated in the same
method as the section 6. We can know the r.m.s. value of the
distribution of the projection of the unit direction vectors
predicted by the BNNs in the same method as the Ref.\cite{key-1}.
From the central limit theorem $\vec{q}$ follows that the
distributions of its three components are Gaussian with
$\sigma=r.m.s./\sqrt{N}$ centered at (0,0,$|\vec{q}|$). Therefore,
the uncertainty on the measurement of the neutrino direction can be
given as the cone around $\vec{q}$ which contains 68.3\% of the
integral of this distribution.

\section{Results}
Fig. 3 shows the distributions of the projections of the
$\hat{X}_{e^+n}$ in the sec. 6 and the $\vec{L}$ predicted by the
method of BNN along the reactor neutrino incoming direction. The
r.m.s. attainable by using BNN is only about 0.41, and less than
that attainable by using the method in the Ref.\cite{key-1}. The
results of the determination of the reactor neutrino incoming
direction using the method in the Ref.\cite{key-1} and the method of
BNN are shown in Table 1. The uncertainty attainable by using the
method in the Ref.\cite{key-1} is 21.1$^\circ$,and the one
attainable by using BNN is 1.0$^\circ$. Fig. 4 shows the
distributions of the projections of the $\hat{X}_{e^+n}$ in the sec.
6 and the $\vec{L}$ predicted by the method of BNN along the
supernova neutrino incoming direction. The r.m.s. attainable by
using BNN is also about 0.35. The results of the determination of
the supernova neutrino incoming direction using the method in the
Ref.\cite{key-1} and the method of BNN are shown in Table 2. The
uncertainty attainable by using the method in the Ref.\cite{key-1}
is 10.7$^\circ$, and the one attainable by using BNN is 0.6$^\circ$.

\par
So compared to the method in Ref.\cite{key-1}, the uncertainty
attainable by using BNN is significantly improved and reduces by a
factor of about 20 (21$^\circ$ compared to 1$^\circ$ in the case of
reactor neutrinos and 11$^\circ$ compared to 0.6$^\circ$ in the case
of supernova neutrinos). And compared to SK, it reduces by a factor
of about 8 (5$^\circ$ compared to 0.6$^\circ$). Why such good
results can be obtained with BNN? First, neutrino directions
obtained with the method in the Ref.\cite{key-1} are used as inputs
to BNN, that is such good results obtained with BNN is on the base
of the results of the method in the Ref.\cite{key-1}; Second, BNN
can extract some unknown information from its inputs and discover
more general relationships in data than traditional statistical
methods; Third, the over-fitting problem can be solved by using
Bayesian methods to control model complexity. So results obtained
with BNN can be much better than that of the method in the
Ref.\cite{key-1}. In a word, the method of BNN can be well applied
to determine neutrino incoming direction in reactor neutrino
experiments and supernova explosion location by scintillator
detectors.

\section{Acknowledgements }

This work is supported by the National Natural Science Foundation
of China (NSFC) under the contract No. 10605014.


\newpage{}

\begin{table}

\caption{Measurement of reactor neutrino direction}
\begin{tabular}{|c|c|c|c|c|}\hline
    & |$\vec{p}$|  or |$\vec{q}$| &$\phi$ &$\theta$ &uncertainty\\\hline
 known neutrino incoming direction &--&60$^\circ$&
 120$^\circ$&--\\\hline
 Direction determined by the method in
 Ref.\cite{key-1}&0.033&42.5$^\circ$&111.4$^\circ$&21.1$^\circ$\\\hline
 Direction determined by BNN
 &0.708&56.7$^\circ$&118.9$^\circ$&1.0$^\circ$\\\hline

\end{tabular}
\end{table}

\begin{table}

\caption{Measurement of supernova neutrino direction}
\begin{tabular}{|c|c|c|c|c|}\hline
    & |$\vec{p}$| or |$\vec{q}$| &$\phi$ &$\theta$ &uncertainty\\\hline
 known neutrino incoming direction &--&60$^\circ$&
 135$^\circ$&--\\\hline
 Direction determined by the method in
 Ref.\cite{key-1}&0.066&61.0$^\circ$&149.2$^\circ$&10.7$^\circ$\\\hline
 Direction determined by BNN
 &0.727&55.8$^\circ$&138.5$^\circ$&0.6$^\circ$\\\hline

\end{tabular}
\end{table}

\newpage{}
\begin{figure}
\includegraphics[width=16cm,height=16cm]{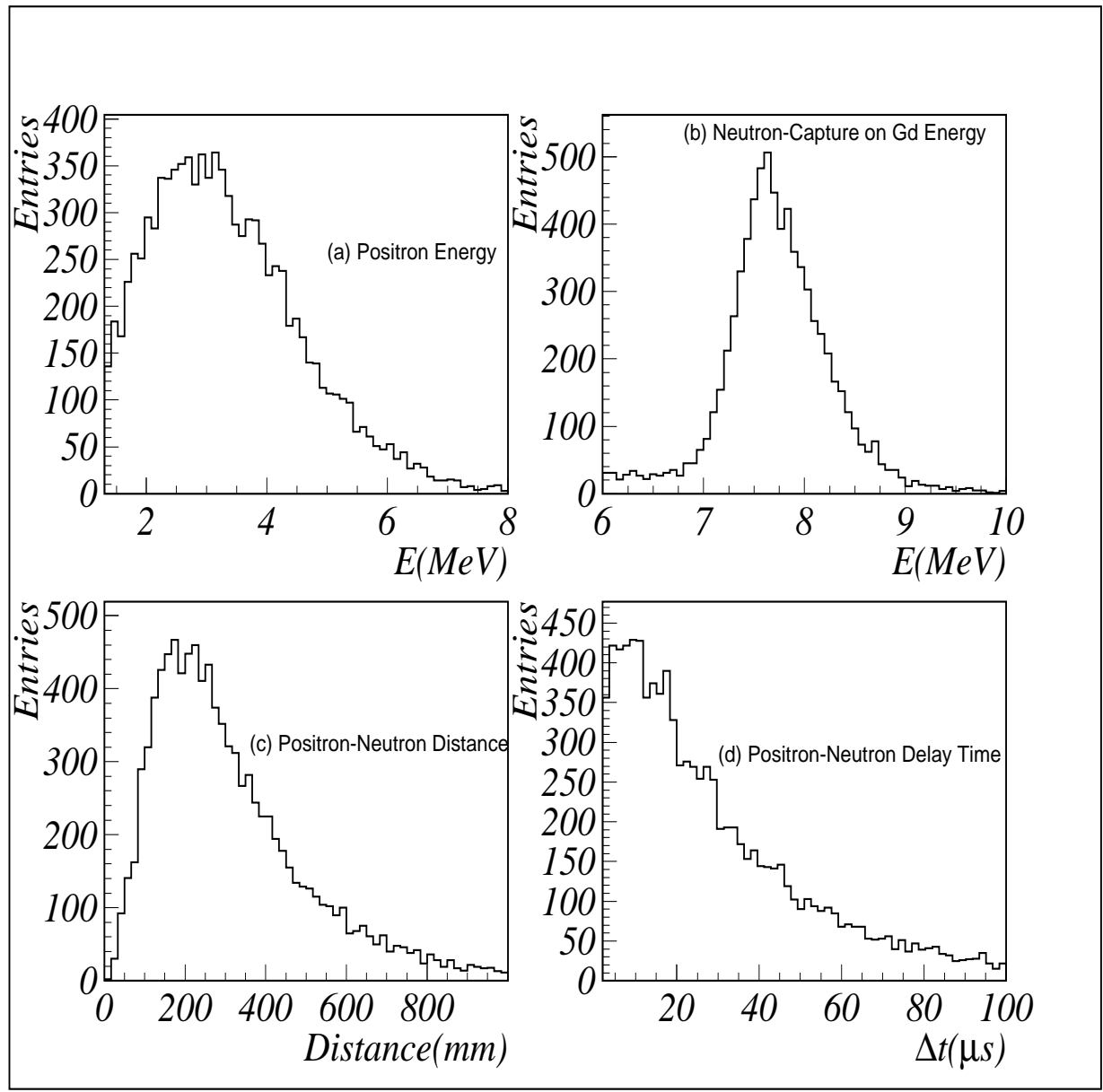}

\caption{The reactor neutrino events for the Monte-Carlo simulation
of the toy detector are uniformly generated throughout Gd-LS region.
(a) is the distribution of the positron energy; (b) is the
distribution of the energy of the neutron captured by Gd; (c) is the
distribution of the distance between the positron and neutron
positions; (d) is the distribution of the delay time of the neutron
signal.}
\end{figure}

\begin{figure}
\includegraphics[width=16cm,height=16cm]{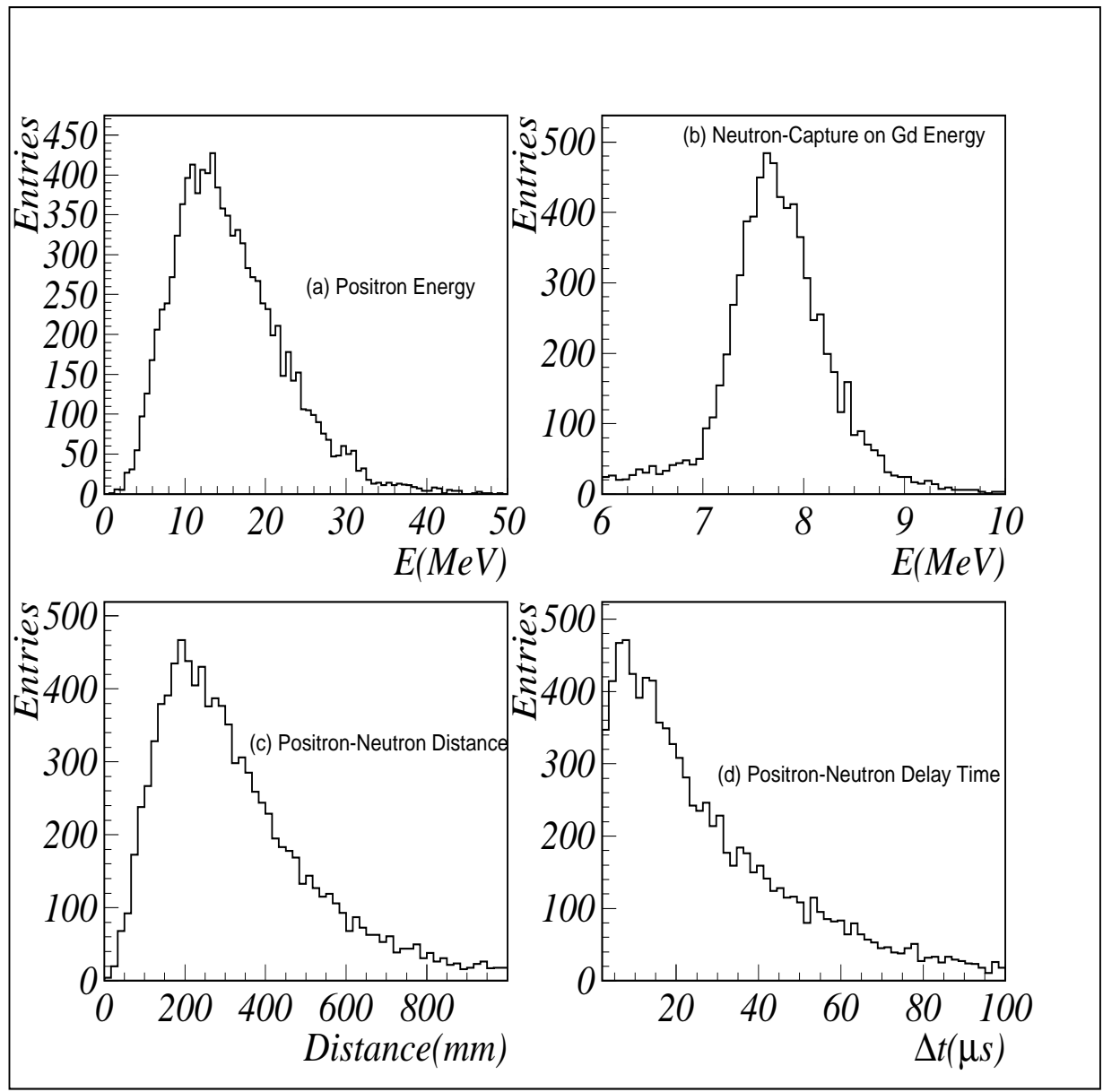}

\caption{The supernova neutrino events for the Monte-Carlo
simulation of a liquid scintillator detector with the same geometry
and the same target as the toy detector in the sec. 3 are uniformly
generated throughout Gd-LS region. (a) is the distribution of the
positron energy; (b) is the distribution of the energy of the
neutron captured by Gd; (c) is the distribution of the distance
between the positron and neutron positions; (d) is the distribution
of the delay time of the neutron signal.}
\end{figure}

\begin{figure}
\includegraphics[width=16cm,height=16cm]{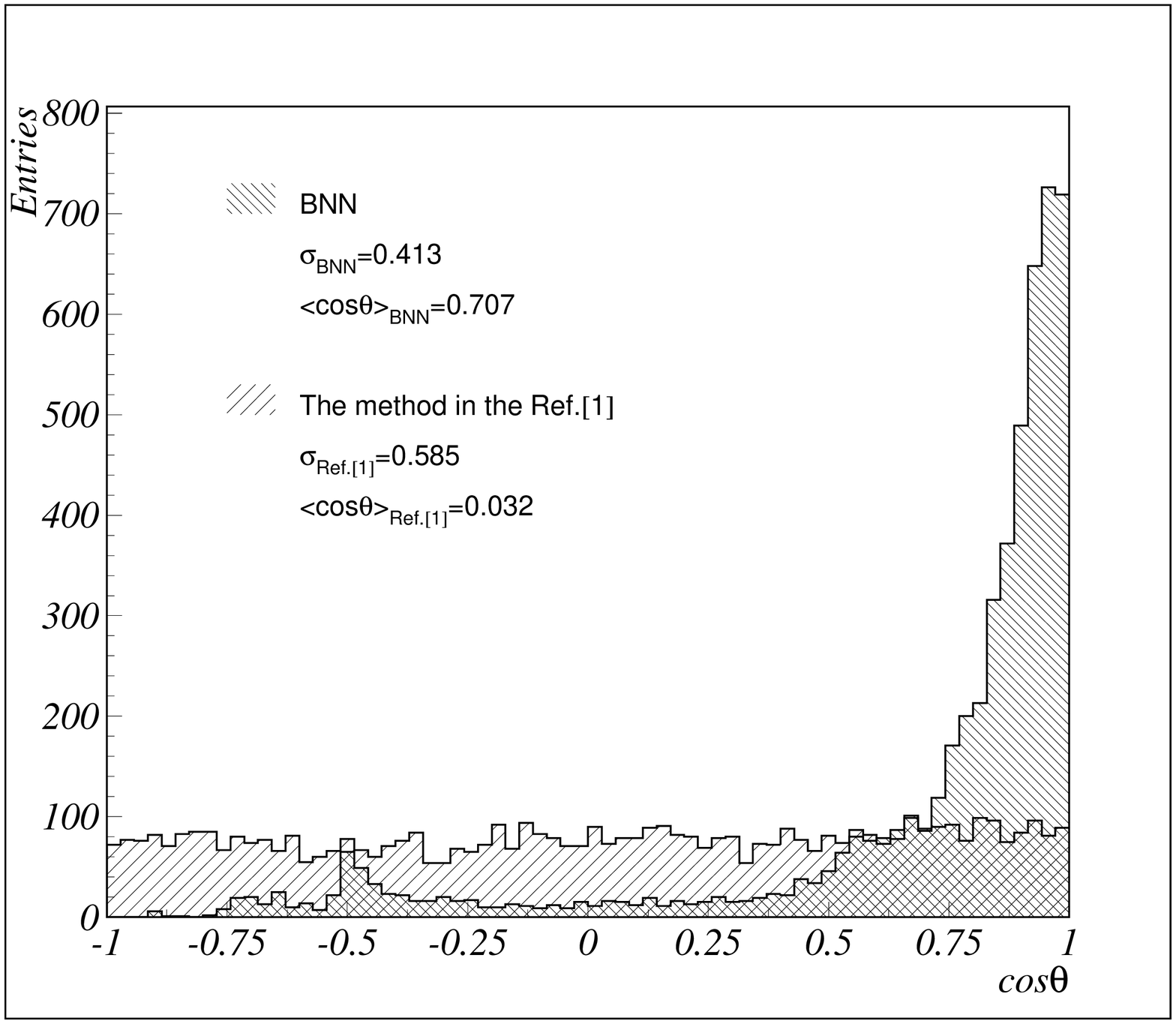}

\caption{The distributions of the projections of the
$\hat{X}_{e^+n}$ in the sec. 6 and the $\vec{L}$ predicted by the
method of BNN along the reactor neutrino incoming direction.}
\end{figure}

\newpage{}

\begin{figure}
\includegraphics[width=16cm,height=16cm]{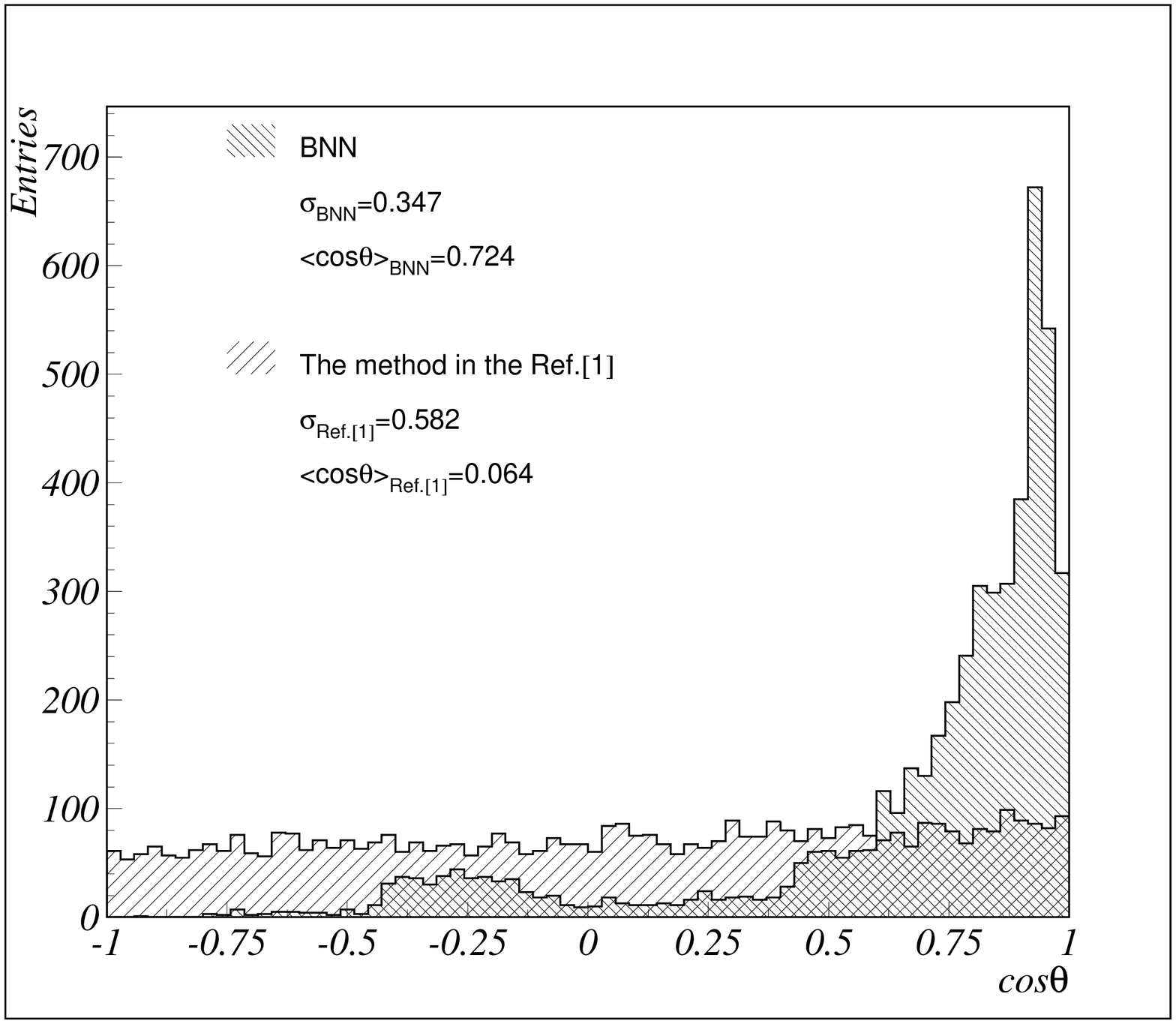}

\caption{The distributions of the projections of the
$\hat{X}_{e^+n}$ in the sec. 6 and the $\vec{L}$ predicted by the
method of BNN along the supernova neutrino incoming direction.}
\end{figure}

\end{document}